\documentclass[journal]{IEEEtran}
\usepackage{cite}
\usepackage{times,amsmath,amssymb,psfrag}
\usepackage{graphicx}
\usepackage{upgreek}
\usepackage{url}
\usepackage{bm}
\usepackage{multicol}
\usepackage{flushend}
\usepackage{diagbox} 
\graphicspath{{./pics/}} \DeclareGraphicsExtensions{.pdf,.jpg,.png}
\usepackage{mdframed}
\usepackage{xcolor}
\usepackage{hyperref}
\usepackage{epsfig}
\usepackage{float}
\usepackage{array}
\usepackage{multirow}
\usepackage{etoolbox}
\usepackage{nomencl}
\usepackage{algorithm}
\usepackage{algorithmicx}
\usepackage{algpseudocode}
\usepackage{enumerate}
\usepackage{amsmath}
\usepackage{amsthm}
\usepackage{soul}
\soulregister\cite7 
\soulregister\citep7 
\soulregister\citet7 
\soulregister\ref7 
\soulregister\pageref7 

\renewcommand\nomgroup[1]{%
  \item[\normalsize\itshape\bfseries
  \ifstrequal{#1}{I}{Parameters/States of Composite Load Model}{%
  \ifstrequal{#1}{P}{Notation for Algorithm}{%
  \ifstrequal{#1}{N}{Notation1 for Algorithm}{%
  \ifstrequal{#1}{X}{Other Smbols}{}}}}]%
  }
\makenomenclature
\begin{document}

\title{Singular Perturbation-based Large-Signal Order Reduction of Microgrids for Stability and Accuracy Synthesis with Control}
%
%
%

\author{Zixiao~Ma,~\IEEEmembership{Member,~IEEE,}
	Zhaoyu~Wang,~\IEEEmembership{Senior Member,~IEEE,}
	Yuxuan~Yuan,~\IEEEmembership{Member,~IEEE,}
	Tianqi~Hong,~\IEEEmembership{Member,~IEEE,}~
	\thanks{Z. Ma, Z. Wang and Y. Yuan are with the Department of Electrical and Computer Engineering, Iowa State University, Ames, IA, 50011 USA (E-mail: zma@iastate.edu; wzy@iastate.edu; yuanyx@iastate.edu). (Corresponding author: Zhaoyu Wang.)}
	\thanks{T. Hong is with Energy System Division, Argonne National Laboratory, Lemont, IL. (Email: thong@anl.gov).}
	
}
\maketitle

\begin{abstract}
With the increasing penetration of distributed energy resources (DERs), it is of vital importance to study the dynamic stability of microgrids (MGs) with external control inputs in the electromagnetic transient (EMT) time scale. This requires detailed models of the underlying control structure of MGs and results in a high-order nonlinear MG control system. Higher-level controller design and stability analysis of such high-order systems are usually intractable and computation-costly. To overcome these challenges, this paper proposes a large-signal order reduction (LSOR) method for MGs with considerations of external control inputs and the detailed dynamics of underlying control levels based on singular perturbation theory (SPT). Specially, we innovatively proposed and strictly proved a general stability and accuracy assessment theorem that allows us to analyze the dynamic stability of a full-order nonlinear system by only leveraging its corresponding reduced-order model (ROM) and boundary layer model (BLM). Moreover, this theorem also theoretically provides a set of conditions under which the developed ROM is accurate. Finally, by embedding such a theorem into the SPT, we propose a novel LSOR approach with guaranteed accuracy and stability analysis equivalence. The proposed LSOR method is generic and can be applied to arbitrary dynamic systems. Multiple case studies are conducted on MG systems to show the effectiveness of the proposed approach. 
\end{abstract}


\markboth{Submitted to IEEE for possible publication. Copyright may be transferred without notice}%
{Shell \MakeLowercase{\textit{et al.}}: Bare Demo of IEEEtran.cls for Journals}

%
\IEEEpeerreviewmaketitle

\section{Introduction}
%
%
%
%
\IEEEPARstart{M}{ICROGRIDS} (MGs) are localized small-scale power systems composed of interconnected loads and distributed energy resources (DERs) in low-voltage and medium-voltage distribution networks. It can be operated in grid-connected and islanded modes \cite{Shahidehpour2010,SHAHIDEHPOUR201221, Ma2021,Zhang2021}. The high penetration of low-inertia DERs makes the dynamic response of MGs different from conventional networks dominated by synchronous machines. This low-inertia characteristic highlights the importance of dynamic modeling, stability analysis, and control studies of MGs in the electromagnetic transient (EMT) time scale \cite{Wang2014,Lu2015,Ma2023}. To precisely capture the comprehensive transient dynamics of MGs in a hierarchical control structure, detailed dynamic models of the lower control levels such as primary and zero-control levels, and the impact of external input from higher control levels such as secondary control, need to be taken into account. However, the high-order nature of these detailed dynamics of the underlying control structures makes it intractable to analyze the stability of MGs with such a complex dynamic model \cite{Li2017,Li2018a,LI2018279,Li2019}. In addition, another critical challenge brought by considering the underlying controllers is the two-time-scale behavior of MGs due to the different evolutionary velocities of different state variables, which leads to a stiff differential equation problem \cite{Wang2018}. In the dynamic simulation of MGs, numerically solving this stiff problem requires extremely small time steps, which results in an unmanageable computational complexity \cite{herath2018deterministic}.

To solve the above problems, model order reduction techniques have been studied and applied to power system analyses. 
In \cite{Purba2019,Shuai2019}, an aggregate equivalent model was developed for the order reduction of MGs by assuming similar inverter dynamics. Kron reduction was adopted to simplify the network of MGs in \cite{Floriduz2019}. In \cite{Wangrui2021}, the authors used a balanced truncation method for DC MGs described by a linear model with inhomogeneous initial conditions. Although these methods can effectively simplify the MG model, the time-scale separation problem aroused by the consideration of underlying control levels for EMT analysis is still not solved. 

Given the inherent two-time-scale property of MGs, singular perturbation theory (SPT) is a suitable technology for this purpose. The SPT is a mathematical framework that focuses on analyzing problems with a parameter, where the solutions of the problem at a specific limiting value of the parameter exhibit distinct characteristics compared to the solutions of the general problem, resulting in a singular limit. It facilitates the separation of the system into a reduced-order model (ROM) that captures the slow states, and a boundary layer model (BLM) that represents the errors between fast and quasi-steady states. It is worth noting that the terms ``slow'' and ``fast'' refer to the transient evolutionary velocity of states in this context. Unlike conventional model reduction methods that simply neglect certain state variables, SPT preserves the characteristics of fast dynamics by integrating them into the ``slow'' states, as advocated by \cite{Khalil2000}. Additionally, SPT has the advantage of converting the original stiff problem into a non-stiff problem, resulting in improved computational efficiency. A spatiotemporal model reduction method of MGs using SPT and Kron reduction was proposed in \cite{Luo2014}, nonetheless, the method is not generic enough. In \cite{Rasheduzzaman2015,vorobev2017high}, a linear SPT was applied to small-signal models of MGs. However, since it used the small-signal model, the result only holds in the neighborhood of a stable equilibrium point. 

The above studies focus on the development of the reduced-order MG modeling, whereas the stability assessment based on the derived ROM is not included. To fill this gap, the system order is reduced for simplifying the stability analysis by neglecting the underlying voltage controller in \cite{Anand2013} at the expense of losing fast dynamics. In \cite{Kabalan2019}, the nonlinear Lyapunov stability of DC/AC inverters with different ROMs was studied. A method for simplifying
the stability assessment was developed and applied to an islanded MG with droop control by using inverter angles in \cite{Simpson-Porco2013}. Nevertheless, it was demonstrated that such a simplification process could affect the accuracy of reduced models in \cite{Mariani2015,Nikolakakos2018}. Moreover, to our best knowledge, the existing studies \textit{do not consider the impact of external inputs} such as power commands and voltage frequency references on MG stability analysis. A typical way is to consider the unforced system by neglecting the inputs to study the internal stability. However, even though the unforced system is stable, a continuous input signal can render the system unstable. In \cite{Christofides1996}, a stability assessment criterion that used the input-to-state stability (ISS) of ROM and global asymptotic stability (GAS) of BLM was proposed to analyze the total stability of the original system. This method is generic for arbitrary singular perturbed systems under certain conditions, nevertheless, the convergence of the error between reduced and original models is not theoretically analyzed, which hinders the accuracy evaluation of reduced models.

To overcome the above challenges, this paper proposes a novel large-signal order reduction (LSOR) strategy for inverter-based MGs with detailed dynamics of the underlying control levels in the EMT time scale. Firstly, a general theorem for analyzing the dynamic stability of the full-order model by only alternatively assessing the stability of its derived ROM and BLM is proposed. A key point is that we consider ISS to quantify the system’s response to external inputs and unify internal and external stability. In particular, by assuming the ROM to be ISS, the unforced ROM to be exponentially stable, and BLM to be uniformly GAS, one can prove that the original system is totally ISS. Then, we develop the conditions that guarantee the accuracy of reduced models for both slow and fast dynamics. Finally, by embedding the proposed stability and accuracy assessment theorem into the large-signal SPT, an improved LSOR algorithm is proposed for MGs. Strict mathematical proof is provided to illustrate that the proposed order reduction technique is generic for arbitrary dynamic systems. The main contributions can be summarized as follows: 
\begin{itemize}
\item  We propose a general theorem that allows us to assess the large-signal stability of MGs with detailed dynamics of underlying controllers in the EMT time scale by only analyzing their ROMs and BLMs. 

\item  A set of accuracy criteria is developed, under which the error between the reduced and original models is bounded and converges as the perturbation coefficients decrease. 

\item The impact of external control input from the higher control level on the above stability and accuracy analyses is studied with strict mathematical proof.

\item  The stability and accuracy assessment synthesis is embedded into the LSOR method  to improve the model accuracy via a feedback mechanism, which automatically tunes the bounds of perturbation coefficients as an index for identifying the slow and fast dynamics. 
\end{itemize}

The rest of the paper is organized as follows. Section \ref{C3} describes the large-signal 
 mathematical model of the studied MG system. Section \ref{C2} introduces the general singular perturbation theory and proposes our stability and accuracy assessment theory. Section \ref{C4} gives the simulation validation of the proposed method. Section \ref{C5} concludes the paper.
\section{Large-Signal Modeling of Inverter-Based MGs}\label{C3}
This section introduces a nonlinear model of the studied MG system with detailed primary and zero-control levels. Depending on the research objectives, control strategies, and operation modes, MGs may have different models. According to \cite{Rasheduzzaman2015}, the transient response velocity of line dynamics is much faster than the slow ones in DERs due to the small line impedance. Moreover, the state equations are fully decoupled between DERs and lines. As a result, the line dynamics can be neglected. Therefore, this section focuses on the modeling of DERs, which are the main dynamic components in an inverter-based MG. 
\begin{figure*}[ht!]
	\centering
	\includegraphics[width=1.8\columnwidth]{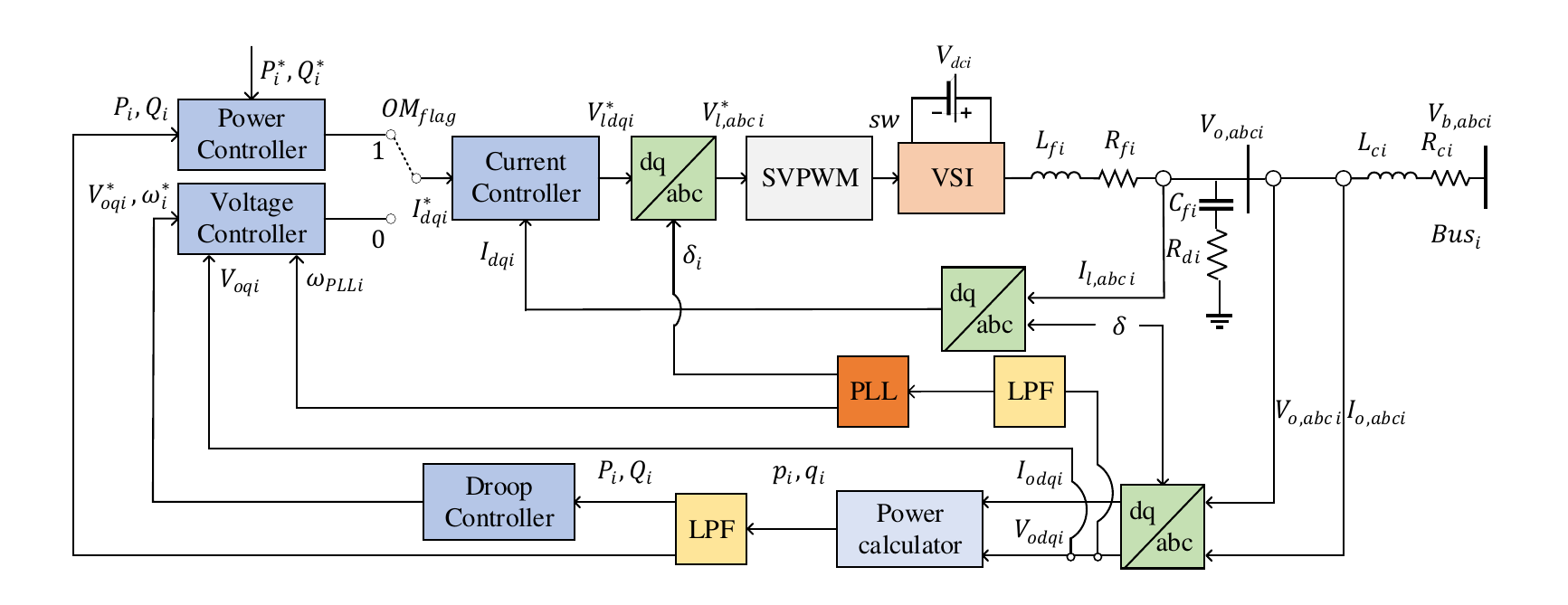}
	\caption{The diagram of VSI-based DER and controller block.}
	\label{diagram_microgrid}
\end{figure*}

A general control diagram of DERs is shown in Fig. \ref{diagram_microgrid}. The model can switch between two subsystems according to the MG operation modes. In grid-tied mode, $\rm OM_{flag}$ switches to $1$, then the voltage source inverter (VSI) is controlled by the power controller and current controller to follow the power command $P^*, Q^*$. The MG bus voltage and system frequency are maintained by the main grid. In islanded mode, $\rm OM_{flag}$ is set to $0$, and the MG voltage and frequency are regulated by the DERs using droop controllers. According to Fig. \ref{diagram_microgrid}, the mathematical model can be derived for each component  where $i=1,\dots, N$ denotes the index of $N$ DERs in the MG.
\subsection{Average Power Calculation}
The generated active and reactive power can be calculated using the transformed output voltage $v_{\rm odq}$ and current $i_{\rm odq}$. Using a low-pass filter (LPF) with the corner frequency $\omega_{\rm c}$, we can obtain the filtered instantaneous powers as follows,
\begin{subequations}\label{pq}
\begin{align}
\dot{P_i}&=-P_i\omega_{{\rm c}i}+1.5\omega_{{\rm c}i}\left( V_{{\rm od}i}I_{{\rm od}i}+V_{{\rm oq}i}I_{{\rm oq}i}\right), \label{P} \\
\dot{Q_i}&=-Q_i\omega_{{\rm c}i}+1.5\omega_{{\rm c}i}\left( V_{{\rm oq}i}I_{{\rm od}i}-V_{{\rm od}i}I_{{\rm oq}i}\right). \label{Q}
\end{align}	
\end{subequations}

\subsection{Phase Lock Loop}
The model of phase lock loop (PLL) is the same as that established in \cite{Rasheduzzaman2015} as follows,
\begin{subequations}\label{pll}
\begin{align}
\dot{V}_{{\rm odf}i}&=\omega_{{\rm cPLL}i}V_{{\rm od}i}-\omega_{{\rm cPLL}i}V_{{\rm odf}i}, \label{vodf} \\
\dot{\varPhi}_{{\rm PLL}i}&=-V_{{\rm odf}i}. \label{varphi}
\end{align}	
\end{subequations}

In grid-tied mode, the inverter output phase is synchronized to the main grid using PLL, therefore the derivative of phase angle $\delta_i$ is set to $\omega_{{\rm PLL}i}$:
\begin{eqnarray}\label{delta_grid}
\dot{\delta}_i=\omega_{{\rm PLL}i}=377-K_{{\rm P,PLL}i}V_{{\rm odf}i}+K_{{\rm I,PLL}i}\varPhi_{{\rm PLL}i}.
\end{eqnarray}

In islanded mode, the phase angle of the first inverter can be arbitrarily set as the reference for the other inverters:
\begin{eqnarray}
\dot{\delta}_i=\omega_{\rm PLL1}-\omega_{{\rm PLL}i}.\label{delta_islanded}
\end{eqnarray}
\subsection{Power Controllers}
In grid-tied mode, the output power of DER is regulated by the power controller using the PI control method. The input references are the commanded real and reactive powers:

\begin{subequations}\label{pc}
	\begin{align}
	\dot{\varPhi}_{{\rm P}i}&=P_{i}-P_{i}^*,\label{pcd1} \\ 
	I_{{\rm lq}i}^*&=K_{{\rm I,P}i}{\varPhi}_{{\rm P}i}+K_{{\rm P,P}i}\dot{\varPhi}_{{\rm P}i}, \label{pcd2} \\
	\dot{\varPhi}_{{\rm Q}i}&=Q_{i}-Q_{i}^*,\label{pcq1} \\ 
	I_{{\rm ld}i}^*&=K_{{\rm I,P}i}{\varPhi}_{{\rm Q}i}+K_{{\rm P,P}i}\dot{\varPhi}_{{\rm Q}i}. \label{pcq2} 
	\end{align}
\end{subequations}
\subsection{Voltage Controllers and Droop Controllers}
In islanded mode, a DER has no reference inputs from the main grid. Therefore, it must generate its only voltage and frequency references using droop controllers as follows,
\begin{subequations}\label{droop}
\begin{align}
\omega_{i}^*&=\omega_{{\rm n}i}-m_iP_i,\label{droop1} \\ 
V_{{\rm oq}i}^*&=V_{{\rm oq,n}i}-n_iQ_i. \label{droop2} 
\end{align}
\end{subequations}

These references will be used as the set points for voltage controllers. Two PI controllers are adopted for the voltage controllers as follows,
\begin{subequations}\label{vc}
\begin{align}
\dot{\varPhi}_{{\rm d}i}&=\omega_{{\rm PLL}i}-\omega_{i}^*,\label{vcd1} \\ 
I_{{\rm ld}i}^*&=K_{{\rm I,V}i}{\varPhi}_{{\rm d}i}+K_{{\rm P,V}i}\dot{\varPhi}_{{\rm d}i}, \label{vcd2} \\
\dot{\varPhi}_{{\rm Q}i}&=V_{{\rm oq}i}^*-V_{{\rm oq}i},\label{vcq1} \\ 
I_{{\rm lq}i}^*&=K_{{\rm I,V}i}{\varPhi}_{{\rm q}i}+K_{{\rm P,V}i}\dot{\varPhi}_{{\rm q}i}. \label{vcq2} 
\end{align}	
\end{subequations}
\subsection{Current Controllers}
The PI controllers are adopted for current controllers. They generate the commanded voltage reference $V_{{\rm ldq}i}^*$ according to the error between the inductor currents reference $I_{{\rm ldq}i}^*$ and its feedback measurements  $I_{{\rm ldq}i}$:
\begin{subequations}\label{cc}
	\begin{align}
	\dot{\Gamma}_{{\rm d}i}&=I_{{\rm ld}i}^*-I_{{\rm ld}i},\label{ccd1} \\ V_{{\rm ld}i}^*&=-\omega_{{\rm n}i}L_{{\rm f}i}I_{{\rm lq}i}+K_{{\rm I,C}i}{\Gamma}_{{\rm d}i}+K_{{\rm P,C}i}\dot{\Gamma}_{{\rm d}i}, \label{ccd2} \\
	\dot{\Gamma}_{{\rm q}i}&=I_{{\rm lq}i}^*-I_{{\rm lq}i},\label{ccq1} \\ V_{{\rm lq}i}^*&=-\omega_{{\rm n}i}L_{{\rm f}i}I_{{\rm ld}i}+K_{{\rm I,C}i}{\Gamma}_{{\rm q}i}+K_{{\rm P,C}i}\dot{\Gamma}_{{\rm q}i}. \label{ccq2} 
	\end{align}	
\end{subequations}
\subsection{LC Filters and Coupling Inductors}
The dynamical models of LC filters and coupling inductors are as follows,
\begin{subequations}\label{lcf}
\begin{align}
\dot{I}_{{\rm ld}i}&=\left(-R_{{\rm f}i}I_{{\rm ld}i}+V_{{\rm ld}i}-V_{{\rm od}i} \right)/L_{{\rm f}i}+\omega_{{\rm n}i}I_{{\rm lq}i} ,\label{ild} \\ 
\dot{I}_{{\rm lq}i}&=\left(-R_{{\rm f}i}I_{{\rm lq}i}+V_{{\rm lq}i}-V_{{\rm oq}i} \right)/L_{{\rm f}i}-\omega_{{\rm n}i}I_{{\rm ld}i} ,\label{ilq} \\
\dot{I}_{{\rm od}i}&=\left(-R_{{\rm c}i}I_{{\rm od}i}+V_{{\rm od}i}-V_{{\rm bd}i} \right)/L_{{\rm c}i}+\omega_{{\rm n}i}I_{{\rm oq}i} ,\label{iod} \\ 
\dot{I}_{{\rm oq}i}&=\left(-R_{{\rm c}i}I_{{\rm oq}i}+V_{{\rm oq}i}-V_{{\rm bq}i} \right)/L_{{\rm c}i}-\omega_{{\rm n}i}I_{{\rm od}i} ,\label{ioq} \\
\dot{V}_{{\rm od}i}&=\left(I_{{\rm ld}i}\!-\!I_{{\rm od}i} \right)/C_{{\rm f}i}\!+\!\omega_{{\rm n}i}V_{{\rm oq}i}+R_{{\rm d}i}(\dot{I}_{{\rm ld}i}-\dot{I}_{{\rm od}i})  ,\label{vod} \\ 
\dot{V}_{{\rm oq}i}&=\left(I_{{\rm lq}i}\!-\!I_{{\rm oq}i} \right)/C_{{\rm f}i}\!-\!\omega_{{\rm n}i}V_{{\rm od}i}+R_{{\rm d}i}(\dot{I}_{{\rm lq}i}-\dot{I}_{{\rm oq}i}) .\label{voq}
\end{align}	
\end{subequations}

In conclusion, when the MG system is operating in grid-tied mode, the mathematical model can be represented by equations (\ref{pq})-(\ref{delta_grid}), (\ref{pc}) and (\ref{cc})-(\ref{lcf}). In islanded mode, the MG model can be represented by equations (\ref{pq})-(\ref{pll}), (\ref{delta_islanded}) and (\ref{droop})-(\ref{lcf}).

\section{Improved LSOR by Embedding Stability and Accuracy Assessment Theorem}\label{C2}
In this section, we propose an improved LSOR method together with stability and accuracy assessment synthesis. Firstly, we briefly present the SPT-based LSOR approach. Then a novel large-signal stability and accuracy assessment theorem with consideration of external control input is proposed. Finally, we improve the LSOR algorithm by embedding the stability and accuracy assessment theorem, so that it can guarantee the accuracy of derived ROM and efficiently evaluate the stability of original models. The proposed LSOR strategy is essentially generic and is suitable for the above MG model introduced in Section \ref{C3}.
\subsection{LSOR Approach using the SPT for MGs}
Due to the two-time-scale property, the dynamics of MGs can be classified as slow and fast dynamics according to the transient velocity. The main idea of SPT is to \textit{freeze} the fast dynamics and degenerate them into static equations. Thus, the ROM can be obtained by substituting the solutions of the static equations into the slow dynamic equations.

Consider an MG system in the following general singular perturbed form,
\begin{subequations}\label{system}
\begin{align}
\dot {\mathbf{x}}(t) &= \mathbf{f}\left( \mathbf{x}(t),\mathbf{z}(t),\mathbf{u}(t),\varepsilon\right),   \label{standardx} \\
\varepsilon\dot {\mathbf{z}}(t) &= \mathbf{\mathbf{g}}\left( \mathbf{x}(t),\mathbf{z}(t),\mathbf{u}(t),\varepsilon\right),  \label{standardz}
\end{align}
\end{subequations}
where $\mathbf{x}\in \mathbb{R}^n$,  $\mathbf{z}\in \mathbb{R}^m$ represent the fast and slow state variables of MGs, respectively, such as voltages and currents; $\mathbf{u} \in \mathbb{R}^p$ denotes the continuous MG inputs such as power commands, or voltage frequency commands. $\varepsilon$ denotes the small parameters in MGs such as capacitances and inductances named as perturbation coefficient. $\mathbf{f}$ and $\mathbf{\mathbf{g}}$ are locally Lipschitz functions on their arguments. For simplicity, we neglect the notation of time-dependency $(t)$ in the rest of this paper. 

Since $\varepsilon$ is small, the fast transient velocity $\dot {\mathbf{z}}=\mathbf{\mathbf{g}}/\varepsilon$ can be much larger than the slow dynamics $\dot{\mathbf{x}}$. To solve this two-time-scale problem, we can set $\varepsilon=0$, then equation (\ref{standardz}) degenerates to the following algebraic equation,
\begin{eqnarray}\label{algebraic}
0 = \mathbf{\mathbf{g}}\left( \mathbf{x},\mathbf{z},\mathbf{u},0\right).
\end{eqnarray}

If equation (\ref{algebraic}) has at least one isolated real root and satisfies the implicit function theory, then for each argument, we have the following closed-form solution, 
\begin{eqnarray}\label{roots}
\mathbf{z} = \mathbf{h}\left( \mathbf{x},\mathbf{u}\right). 
\end{eqnarray}

Substitute equation (\ref{roots}) into equation (\ref{standardx}) and let $\varepsilon=0$, we have a quasi-steady-state (QSS) model,
\begin{eqnarray}\label{quasi}
\dot {\mathbf{x}} = \mathbf{f}\left( \mathbf{x},\mathbf{h}\left(\mathbf{x},\mathbf{u}\right),\mathbf{u},0\right). 
\end{eqnarray}

Note that the order of the QSS system {\eqref{quasi} drops from $n+m$ to $n$}. The inherent two-time-scale property can be described by introducing the BLM. Define a fast time scale variable $\tau= t/\varepsilon$, and a new coordinate $\mathbf{y}=\mathbf{z}-\mathbf{h}(\mathbf{x},\mathbf{u})$. In this new coordinate, equation (\ref{standardz}) is rewritten as 
\begin{align}\label{dy}
\frac{d\mathbf{y}}{d\tau} = \;&\mathbf{g}\left( \mathbf{x},\mathbf{y}+\mathbf{h}\left(\mathbf{x},\mathbf{u}\right),\mathbf{u},\varepsilon\right)\nonumber\\
&-\varepsilon \left[\frac{\partial \mathbf{h}}{\partial \mathbf{x}} \mathbf{f}\left(\mathbf{x},\mathbf{y}+\mathbf{h}\left(\mathbf{x},\mathbf{u}\right),\mathbf{u},\varepsilon \right) + \frac{\partial \mathbf{h}}{\partial \mathbf{u}}\dot{\mathbf{u}}\right]. 
\end{align}
Let $\varepsilon=0$, we obtain the BLM as follows, 
\begin{eqnarray}\label{blm}
\frac{d\mathbf{y}}{d\tau} = \mathbf{g}\left( \mathbf{x},\mathbf{y}+\mathbf{h}\left(\mathbf{x},\mathbf{u}\right),\mathbf{u},0\right). 
\end{eqnarray}
\subsection{Stability and Accuracy Assessment Theorem}\label{C2.B}
In this subsection, we propose a criterion to assess the stability of the original system and the accuracy of ROM and BLM. Considering the impact of external inputs on the stability of MGs, we define the ISS as follows.

\emph{Definition 1 (ISS):}
Consider such a nonlinear system
\begin{eqnarray}\label{ns}
	\dot{\mathbf{x}}=\tilde{\mathbf{f}}\left(\mathbf{x},v_1,v_2 \right) 
\end{eqnarray}
where $\mathbf{x}\in \mathbb{R}^n$ is the state vector, $v_1\in \mathbb{R}^m$, $v_2\in \mathbb{R}^p$ are input vectors, and $\tilde{\mathbf{f}}$ is locally Lipschitz on $\mathbb{R}^n\times \mathbb{R}^m \times \mathbb{R}^p$. The system (\ref{ns}) is ISS with Lyapunov gains $\alpha_{v_1}$ and $\alpha_{v_2}$ of class kappa ($\mathcal{K}$), if there exists a class kappa-ell ($\mathcal {KL}$) function $\beta$ such that for $\mathbf{x}\left(0 \right)\in\mathbb{R}^n$ and bounded inputs $v_1$, $v_2$, the solution of (\ref{ns}) exists and satisfies
\begin{eqnarray}\label{ISS}
\left\|\mathbf{x}(t)\right\|\leqslant\beta\left(\left\|\mathbf{x}(0) \right\|,t  \right)   +\alpha_{v_1}\left(\left\|v_1 \right\|  \right)  +\alpha_{v_2}\left(\left\|v_2 \right\|  \right).
\end{eqnarray}

The above definition indicates that an MG system is ISS when all the trajectories are bounded by some functions of the input magnitudes. Then we give the following three assumptions which are the sufficient conditions for the theorem.

\emph{Assumption 1 (Growth conditions):} The functions $\mathbf{f}$, $\mathbf{g}$, and their first partial derivatives are continuous and bounded with respect to $(\mathbf{x},\mathbf{z},\mathbf{u},\varepsilon)$; $\mathbf{h}$ and its first partial derivatives $\partial \mathbf{h}/\partial \mathbf{x}$, $\partial \mathbf{h}/\partial \mathbf{u}$ is locally Lipschitz; and the Jacobian $\partial \mathbf{g}/\partial \mathbf{z}$ has bounded first partial derivatives with respect to its arguments. 

\emph{Assumption 2 (Stability of ROM):} The ROM (\ref{quasi}) is ISS with Lyapunov gain $\hat{\alpha}_x$, and its unforced system has an exponentially stable equilibrium at the origin.

\emph{Assumption 3 (Stability of BLM):} The origin of the BLM (\ref{blm}) is a GAS equilibrium, uniformly in $\mathbf{x}\in \mathbb{R}^n$, $\mathbf{u}\in \mathbb{R}^p$.

    \emph{Remark 1:} The conditions in Assumption 1 are commonly satisfied for most MGs \cite{Mariani2015}. Inspired by  \cite{Christofides1996}, we propose the stability and accuracy assessment of MGs as the following theorem. Note that the conditions, results and proof of our theorem and \cite{Christofides1996} are different. In \cite{Christofides1996}, only the stability of the original system is proved, nonetheless, the accuracy of the ROM and BLM is not analyzed, which is of vital importance to make sure that the derived reduced-order model is correct. However, the addition of accuracy analysis arouses new challenges in the proof which cannot be solved by directly using \cite{Christofides1996}. Therefore, we add a constraint condition on the transient speed in Assumption 2 and propose a new proving method for our theorem.

\emph{Theorem 1:}
If the MGs system (\ref{system}), its ROM (\ref{quasi}) and the BLM (\ref{blm}) satisfy the Assumptions 1-3, then for each pair of $(\mu,\xi)$, there exists a positive constant $\varepsilon^*$, such that for all  $t\in [0,\infty)$, $\max\left\lbrace\left\|\mathbf{x}(0) \right\|,\left\|\mathbf{y}(0) \right\|,\left\| \mathbf{u} \right\|,\left\| \dot{\mathbf{u}} \right\|  \right\rbrace \leqslant \mu$, and $\varepsilon \in (0,\varepsilon^*]$ the errors between the solutions of the original MGs system (\ref{system}) and its ROM (\ref{quasi}) and BLM (\ref{blm}) satisfy
\begin{align}
\|\mathbf{x}(t,\varepsilon)-\hat{\mathbf{x}}(t)\| =O( \varepsilon) , \label{slow} \\
\|\mathbf{z}(t,\varepsilon)-\mathbf{h}(\hat{\mathbf{x}}(t),\mathbf{u}(t))-\hat{\mathbf{y}}(t/\varepsilon)\| =O( \varepsilon), \label{fast}
\end{align}
where $\hat{\mathbf{x}}(t)$ and $\hat{\mathbf{y}}(\tau)$ are the solutions of ROM (\ref{quasi}) and BLM (\ref{blm}), respectively. $\left\| \mathbf{x}-\hat{\mathbf{x}}\right\| =O(\varepsilon)$ means that $\left\| \mathbf{x}-\hat{\mathbf{x}}\right\|\leqslant k\left\|\varepsilon\right\|$ for some positive constant k. Furthermore, for any given $T>0$, there exists a positive constant $\varepsilon^{**}\leqslant\varepsilon^*$ such that for $t\in\left[ T,\infty\right) $ and $\varepsilon<\varepsilon^{**}$, it follows uniformly that
\begin{eqnarray}\label{morefast}
\|\mathbf{z}(t,\varepsilon)-\mathbf{h}(\hat{\mathbf{x}}(t),\mathbf{u}(t))\|=O( \varepsilon).
\end{eqnarray}

Moreover, there exist class $\mathcal {KL}$ functions $\beta_x$, $\beta_y$, a Lyapunov gain $\alpha_x$ of class $\mathcal{K}$ and positive constants $\xi$, such that the solutions of the original MGs system (\ref{standardx}) and (\ref{dy}) exist and satisfy
\begin{align}
\|\mathbf{x}(t,\varepsilon)\| &\leqslant \beta_x\left(\left\|\mathbf{x}(0) \right\|,t  \right)+\alpha_x\left( \left\|\mathbf{u} \right\| \right) + \xi , \label{stab_x} \\
\|\mathbf{y}(t,\varepsilon)\| &\leqslant\beta_y\left(\left\|\mathbf{y}(0) \right\|,\frac{t}{\varepsilon}\right)  + \xi. \label{stab_y}
\end{align}

\emph{Remark 2:} Theorem 1 indicates large-signal stability by observing that $\mu$ can be arbitrarily large. This is more comprehensive than the small-signal stability studied in \cite{Rasheduzzaman2015}. Moreover, the errors between the solutions of reduced and original MGs should be small and bounded to guarantee accuracy. (\ref{slow}) and (\ref{fast}) show that for sufficiently small $\varepsilon$, these errors tend to be zero. Equation (\ref{morefast}) means that for small enough $\varepsilon$, the solution $\hat{\mathbf{y}}$ of the BLM decays to zero exponentially fast in time $T$, so that the fast solutions can be estimated by only QSS solutions $\mathbf{h}(t,\bar{\mathbf{x}}(t))$ after time $T$. 

\emph{Remark 3:} According to the theorem, if the ROM is ISS and BLM is GAS, then the original system is stable as shown in (\ref{stab_x}) and (\ref{stab_y}). Moreover, in real physical systems, one challenge of SPT is how to identify the slow and fast dynamic states. A commonly-used approach is the knowledge discover-based method that relies on expert knowledge for specific domains. For example, in MGs, some small parasitic parameters such as capacitances, inductances, and small time constants, can be selected as the perturbation coefficients $\varepsilon$. The states with respect to these small $\varepsilon$ are identified as fast states. This conventional empirical identification method falls short of efficiency and accuracy. Therefore, we propose a more efficient and accurate method to identify the slow/fast dynamics by finding the bound of $\varepsilon$ in the following proof.

\begin{proof}
Firstly, the stability of $\mathbf{y}$ (\ref{stab_y}) has already been proved by \cite{Christofides1996} using the converse theorem, causality, signal truncation, and Assumption 3. 

Then, we directly use its result to prove the accuracy of the reduced model (\ref{slow})-(\ref{morefast}). Define the error between solutions of reduced and original slow dynamics as $\mathbf{E}_x=\mathbf{x}-\hat{\mathbf{x}}$. When $\varepsilon=0$,  $\mathbf{y}=\mathbf{z}-\mathbf{h}(\mathbf{x},\mathbf{u})=0$. Then, we have
\begin{eqnarray}\label{dotE1}
\dot{\mathbf{E}}_x=\mathbf{f}(\mathbf{E}_x,0,\mathbf{u},0)+\Delta \mathbf{f},
\end{eqnarray}
where $\Delta \mathbf{f}\!=\!\left[\mathbf{f}(\hat{\mathbf{x}}\!+\!\mathbf{E}_x,0,\mathbf{u},0)\!-\!\mathbf{f}(\hat{\mathbf{x}},0,\mathbf{u},0)\!-\!\mathbf{f}(\mathbf{E}_x,0,\mathbf{u},0) \right]\\+\mathbf{f}(\mathbf{x},\mathbf{y},\mathbf{u},\varepsilon)-\mathbf{f}(\mathbf{x},0,\mathbf{u},0) $. According to Assumption 1, it follows that
\begin{align}\label{deltaf}
\left\|\Delta \mathbf{f} \right\| \leqslant& \ell_1\left\|\mathbf{E}_x \right\|^2 +\ell_2\left\|\mathbf{E}_x \right\|\left\|\hat{\mathbf{x}} \right\|\nonumber\\&+\ell_3 \beta_y\left(\left\|\mathbf{y}(0) \right\|,\frac{t}{\varepsilon}\right) 
  + \ell_3\xi+\ell_4\varepsilon,
\end{align}
for some positive constants $\ell_1,\ell_2,\ell_3,\ell_4$. The last term in system (\ref{dotE1}) can be viewed as a perturbation of 
\begin{eqnarray}\label{dotE2}
\dot{\mathbf{E}}_x=\mathbf{f}(\mathbf{E}_x,0,\mathbf{u},0).
\end{eqnarray}
Since the origin of the system (\ref{dotE2}) is exponentially stable with $\mathbf{u}=0$, using the converse theorem, there exist a Lyapunov function $V_{2}(\mathbf{E}_x)$, and positive constants $c_1,c_2,c_3,c_4$, for which it follows that
\begin{align}
c_1\left\|\mathbf{E}_x \right\|^2    \leqslant V_{2}(\mathbf{E}_x)&\leqslant c_2\left\|\mathbf{E}_x \right\|^2  , \label{converse1} \\
\frac{\partial V_{2}}{\partial \mathbf{E}_x}\mathbf{f}(\mathbf{E}_x,0,\mathbf{u},0)\leqslant&-c_3\left\|\mathbf{E}_x \right\|^2 , \label{converse2}\\
\left\|\frac{\partial V_{2}}{\partial \mathbf{E}_x} \right\| \leqslant c_4 &\left\|\mathbf{E}_x \right\|. \label{converse3}
\end{align}
Using (\ref{stab_y}), (\ref{deltaf}) and (\ref{converse1})-(\ref{converse3}), the Lyapunov function of (\ref{dotE2}) along the trajectory of (\ref{dotE1}) satisfies
\begin{align}
\dot{V}_2=&\frac{\partial V_{2}}{\partial \mathbf{E}_x}\mathbf{f}(\mathbf{E}_x,0,\mathbf{u},0) + \frac{\partial V_{2}}{\partial \mathbf{E}_x}\Delta \mathbf{f} \nonumber \\
\leqslant& -c_3\left\|\mathbf{E}_x \right\|^2+c_4 \left\|\mathbf{E}_x \right\|\left[  \ell_1\left\|\mathbf{E}_x \right\|^2 +\ell_2\left\|\mathbf{E}_x \right\|\left\|\hat{\mathbf{x}} \right\|\right. \nonumber\\
&\left. +\ell_3 \beta_y\left(\left\|\mathbf{y}(0) \right\|,\frac{t}{\varepsilon}\right) + \ell_3\xi+\ell_4\varepsilon\right].
\end{align}
For $\left\| \mathbf{E}_x\right\| \leqslant c_3/(2c_4\ell_1)$, using Assumption 2, it follows that
\begin{align}
\dot{V}_2\leqslant& -2\left\lbrace c_3-c_4\ell_1 \left[\hat{\beta}_x\left(\left\|\hat{\mathbf{x}}(0) \right\|,t  \right)+\hat{\alpha}_x\left( \left\|\mathbf{u} \right\| \right) \right]  \right\rbrace V_2\nonumber\\
&+2\left[ \ell_3 \varepsilon+\ell_3\xi+\ell_4 \beta_y\left(\left\|\mathbf{y}(0) \right\|,\frac{t}{\varepsilon}\right)\right]  \sqrt{V_2}\nonumber\\
\leqslant& -2\left\lbrace \ell_a-\ell_b  \hat{\beta}_x\left(\left\|\hat{\mathbf{x}}(0) \right\|,t  \right)  \right\rbrace V_2\nonumber\\
&+2\left[ \ell_c \varepsilon+\ell_d \beta_y\left(\left\|\mathbf{y}(0) \right\|,\frac{t}{\varepsilon}\right)\right]  \sqrt{V_2},
\end{align}
where $0<\ell_a \leqslant c_3-c_4\ell_1\hat{\alpha}_x\left(sup \left\|\mathbf{u} \right\|  \right) $, $\ell_c\geqslant\ell_3(1+\xi/\varepsilon)>0$, and $\ell_b, \ell_d>0$. Using the comparison lemma, we have 
\begin{align}
W_2(t)\leqslant& \phi (t,0)W_2(0)\nonumber\\
&+\int_{0}^{t}{\phi(t,s)\left[  \ell_c \varepsilon+\ell_d \beta_y\left(\left\|\mathbf{y}(0) \right\|,\frac{t}{\varepsilon}\right)\right] }ds,
\end{align}
where $W_2=\sqrt{V_2}$ and 
\begin{align}
\left|\phi(t,s) \right|\leqslant \ell_e e^{-\ell_f t},\;\;\;for\;\;\ell_e,\ell_f>0.
\end{align}
Because
\begin{align}
\int_{0}^{t}e^{-\ell_f t}\beta_y\left(\left\|\mathbf{y}(0) \right\|,\frac{t}{\varepsilon}\right)ds=O(\varepsilon),
\end{align}
it can be verified that $W_2(t)=O(\varepsilon)$. Then it follows that $\mathbf{E}_x(t,\varepsilon)=O(\varepsilon)$, and this means that (\ref{slow}) holds.

Since we have already verified that (\ref{stab_y}) holds in the first step, then by Assumption 3, it follows that
\begin{align}
\!\!\!\mathbf{E_y}(t,\varepsilon)&=\left\| \mathbf{z}(t,\varepsilon)-\mathbf{h}(\hat{\mathbf{x}}(t,\varepsilon),\mathbf{u}(t))-\hat{\mathbf{y}}(t/\varepsilon)\right\|\nonumber\\
&=\left\| \mathbf{y}(t,\varepsilon)-\hat{\mathbf{y}}(t/\varepsilon) \right\|\leqslant \left\|\mathbf{y}(t,\varepsilon) \right\|+\left\| \hat{\mathbf{y}}(t/\varepsilon)\right\|   \\
&\leqslant\beta_y\left(\left\|\mathbf{y}(0) \right\|,t/\varepsilon\right)  \!+ \!\alpha_y(\varepsilon) \!+\! \hat{\beta}_y\left(\left\|\hat{\mathbf{y}}(0) \right\|,t/\varepsilon\right)=O(\varepsilon)\nonumber
\end{align}
for given initial points and all $t\geqslant0$. This proves (\ref{fast}). According to Assumption 3, $\hat{\mathbf{y}}(t/\varepsilon)=\hat{\beta}_y\left(\left\|\mathbf{y}(0) \right\|,t/\varepsilon\right)\to0$ as $\varepsilon\to0$. Thus, the term $\hat{\mathbf{y}}(t/\varepsilon)=O(\varepsilon)$ for all $t\geqslant T>0$ if $\varepsilon$ is small enough to satisfy 
\begin{align}\label{small}
\hat{\beta}_y(\left\|\mathbf{y}(0) \right\|,t/\varepsilon )\leqslant k\varepsilon
\end{align}
Let $\varepsilon^{**}$ and $T$ denote a solution of (\ref{small}) with equal sign. Subsequently, (\ref{morefast}) holds for all $\varepsilon\leqslant\varepsilon^{**}$ uniformly on $[T,\infty)$.

Finally, we prove the ISS of original slow dynamics. Since
\begin{align}
\|\mathbf{x}(t,\varepsilon)\|  -\| \hat{\mathbf{x}}(t)\|\leqslant\|\mathbf{x}(t,\varepsilon)-\hat{\mathbf{x}}(t)\| =O( \varepsilon),
\end{align}
there exist some class $\mathcal {KL}$ function $\beta_x$, class $\mathcal {K}$ function $\alpha$ and a small positive constant $\varepsilon_3$, such that the solution of (\ref{standardx}) exists for all $t\geqslant0$ and $\varepsilon\leqslant \varepsilon^*:=\min\left\lbrace \varepsilon_1,\varepsilon_2,\varepsilon_3\right\rbrace $ satisfying
\begin{align}
\|\mathbf{x}(t,\varepsilon)\|&\leqslant  \| \hat{\mathbf{x}}(t)\|+O( \varepsilon)\nonumber\\
&\leqslant \hat{\beta}_x\left(\left\|\hat{\mathbf{x}}(0) \right\|,t  \right)+\hat{\alpha}_x\left( \left\|\mathbf{u} \right\| \right) +O(\varepsilon)\nonumber\\
&\leqslant \beta_x\left(\left\|\mathbf{x}(0) \right\|,t  \right)+\alpha_x\left( \left\|\mathbf{u} \right\| \right) +\xi.
\end{align}
This completes the proof of (\ref{stab_x}).
\end{proof}
\subsection{Stability and Accuracy Assessment Embedded LSOR}\label{C2.C}
\begin{figure}[tb!]
	\centering
	\includegraphics[width=0.95\columnwidth]{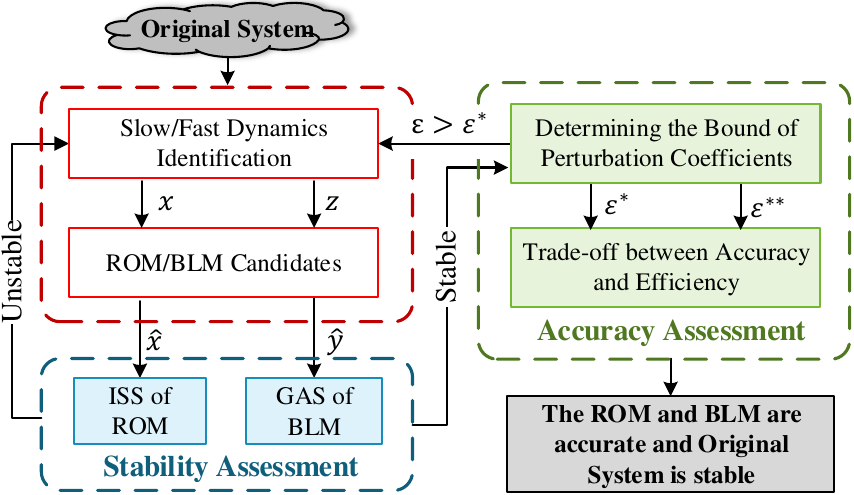}
	\caption{The diagram of stability and accuracy assessment embedded LSOR.}
	\label{algorithm}
\end{figure}

This subsection develops a novel LSOR method by embedding the above theorem. The overall flowchart is shown in Fig. \ref{algorithm} and the detailed algorithm is proposed in \textit{Alogrithm 1}.

This algorithm is designed for MGs with two-time-scale properties, however, no basic assumptions of the MGs are required. Therefore, the proposed method can be applied to arbitrary dynamic systems.

\section{Case Study}\label{C4}
\subsection{Simulation Setup}
\begin{figure}[tb!]
	\centering
	\includegraphics[width=0.95\columnwidth]{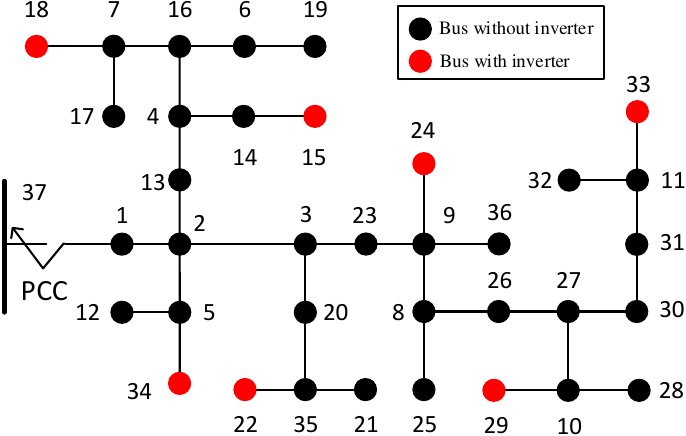}
	\caption{The diagram of modified IEEE-37 bus system.}
	\label{IEEE37}
\end{figure}

The proposed method is tested on a modified IEEE-37 bus MG, which can be operated in grid-tied or islanded modes as shown in Fig. \ref{IEEE37}. According to \cite{Luo2014}, seven inverters are connected to buses 15, 18, 22, 24, 29, 33, and 34. When PCC is closed, the MG is operated in grid-tied mode. Otherwise, it is operated in islanded mode. 

We first let the MG be operated in grid-tied mode. In order to analyze the detailed dynamic properties of both slow and fast dynamics as well as compare our method with the small-signal order reduction approach, a single bus of interest (bus 34) is chosen to show its dynamic responses after power command (input) changes for clearance. Then, a simulation is conducted in islanded mode to show the dynamic responses of multiple buses with DERs when a large load sudden change is given to verify its effectiveness against large disturbances. The detailed load and line parameter settings can be found in \cite{Luo2014}.
\begin{algorithm}
\caption{Stability/Accuracy Assessment Embedded LSOR}\label{alg:LSOR}
\begin{algorithmic}[1]
\State { Choose the smaller parameters dominating the transient velocity as $\varepsilon$. The states with respect to $\varepsilon$ are identified as fast states, while the others as slow states.}
\Procedure{ROM and BLM derivation}{}
    \State {Let $\varepsilon=0$, solve the algebraic equation (\ref{algebraic}) to obtain the isolated QSS solutions $\mathbf{z} = \mathbf{h}\left({\mathbf{x}},\mathbf{u}\right)$}
    \State {Substitute $\mathbf{z}$ into (\ref{standardx}), obtaining the ROM (\ref{quasi})}
    \State {Derive the BLM using equation (\ref{blm}).}
    \EndProcedure
\Procedure{Stability Assessment}{}
    \If {\textit{ Assumption 2 and 3 are satisfied}}
        \State {Go to next procedure}
    \Else   \State {Return to Step 1 to re-identify slow/fast dynamics.}
    \EndIf
\EndProcedure
\Procedure{Calculate the Bound of $\varepsilon$}{}
    \State {Calculate $\varepsilon^*=\min\left\lbrace\varepsilon_1,\varepsilon_2,\varepsilon_3\right\rbrace$ according to proof.}
    \State {Calculate $\varepsilon^{**}$ by solving equation (\ref{small}) with equal sign.}
\EndProcedure
\Procedure{Accuracy Assessment}{}
    \If {\texttt{$ \varepsilon\leqslant\varepsilon^{*}$ }}
        \If{$\varepsilon\leqslant\varepsilon^{**}$}
        \State {$\mathbf{z}=\mathbf{h}(\hat{\mathbf{x}},\mathbf{u})$ is the solution of fast dynamics}
        \Else   \State Use $\mathbf{z}=\mathbf{h}(\hat{\mathbf{x}},\mathbf{u})+\hat{\mathbf{y}}$ by solving the BLM (\ref{blm}).
        \EndIf
    \Else  \State{Return to Step 1 to re-identify slow/fast dynamics}
    \EndIf
\EndProcedure
\end{algorithmic}
\end{algorithm}
\subsection{Performance in Grid-tied Mode and Comparison With Small-Signal ROM}
According to the \textit{Algorithm \ref{alg:LSOR}} in section \ref{C2.C}, we first identify the slow and fast dynamics by finding the $\varepsilon$. Considering the MG model in grid-tied mode, the derivative term can be rewritten as 
\begin{multline}\label{dynamics_case1}
\left[ \frac{1}{\omega _{\rm c}}\dot{P}_i, \frac{1}{\omega _{\rm c}}\dot{Q}_i,  \dot \varPhi_{{\rm PLL}i}, \dot{\delta}_i, \frac{K_{{\rm P,P}i}}{K_{{\rm I,P}i}}{\dot \varPhi _{{\rm P}i}}, \frac{{{K_{{\rm P,P}i}}}}{{{K_{{\rm I,P}i}}}}{\dot \varPhi_{{\rm Q}i}}, \right. \\\left.\frac{K_{{\rm P,C}i}}{K_{{\rm I,C}i}}\dot{\Gamma}_{{\rm d}i},\frac{K_{{\rm P,C}i}}{K_{{\rm I,C}i}}{\dot{\Gamma}_{{\rm q}i}}, \frac{1}{\omega _{{\rm c,PLL}i}}{\dot V_{{\rm od,f}i}},{L_{{\rm f}i}} {\dot I_{{\rm ld}i}}, \right. \\\left. {L_{{\rm f}i}} {\dot I_{{\rm lq}i}}, {L_{{\rm c}i}} {\dot I_{{\rm od}i}},{L_{{\rm c}i}}{\dot I_{{\rm oq}i}}, {C_{{\rm f}i}}{\dot V_{{\rm od}i}},{C_{{\rm f}i}}{\dot V_{{\rm oq}i}}\right]^T
\end{multline}

Substituting the parameters in \cite{Luo2014} into the vector (\ref{dynamics_case1}), it can be seen that the magnitudes of different parameters vary significantly, which is caused by the two-time-scale property of the system. The smaller parameters are selected as perturbation coefficients $\varepsilon$, which are utilized to classify the slow and fast states in this system:
\begin{align}
	\mathbf{x}_1&=\left[P_i,Q_i, \varPhi _{{\rm PLL}i},  \delta_i, \varPhi _{{\rm P}i},\varPhi _{{\rm Q}i}, \Gamma _{{\rm d}i},\Gamma _{{\rm q}i},\right]^T,\label{x1}\\
	\mathbf{z}_1&=\left[ V_{{\rm odf}i},  I_{{\rm ld}i},  I_{{\rm lq}i},  I_{{\rm od}i}, I_{{\rm oq}i}, V_{{\rm od}i}, V_{{\rm oq}i}\right]^T\label{z1}. 
\end{align}

We first set $\varepsilon$ to 0 and calculate the QSS solution $\mathbf{z}_1 = \mathbf{h}\left(\mathbf{x}_1,\mathbf{u}_1\right)$ by solving the algebraic equation with respect to the fast dynamics (\ref{z1}). Then the ROM is obtained by substituting $\mathbf{z}_1$ into the slow dynamic equations with respect to (\ref{x1}). Comparing the numbers of state variables in equation (\ref{dynamics_case1}) and (\ref{x1}), the order of the original model is reduced to $53.33 \%$. Then we derive the BLM using equation (\ref{blm}). Once the ROM and BLM are obtained, we use the conventional ISS and GAS judging theorems in \cite{Khalil2000} to evaluate their stability of them. Specially, the unforced nonlinear ROM is exponentially stable by checking that its linearized system matrix has eigenvalues with strictly negative real parts. It can be verified that the assumptions are satisfied. Based on this result, we are inclined to anticipate the stability of the original system.  

To ensure this, we still need to theoretically verify the accuracy of the ROM and BLM. Following the technique in the proof, we can calculate the boundary of $\varepsilon$ as $\varepsilon^*=\min\left\lbrace \varepsilon_1,\varepsilon_2,\varepsilon_3\right\rbrace=7.92\times10^{-3}$. Note that $\max\left\lbrace\varepsilon \right\rbrace =3.9\times10^{-3}<7.92\times10^{-3}=\varepsilon^*$. Therefore, we can conclude that this MGs system is stable and we can use the solutions of its ROM $\hat{\mathbf{x}}$ and $\mathbf{z}=\mathbf{h}(\hat{\mathbf{x}},\mathbf{u})+\hat{\mathbf{y}}$ to accurately represent its real dynamic responses. Furthermore, given $T=0.43$ s, we can find a $\varepsilon^{**}$ satisfying $\max\left\lbrace\varepsilon \right\rbrace<\varepsilon^{**}=4.2\times10^{-3}$, which indicates that the term $\hat{\mathbf{y}}$ will be $O(\varepsilon)$ after $0.43$ s. Here, a trade-off exists between accuracy and efficiency. When the accuracy is prior, one can choose $\mathbf{z}=\mathbf{h}(\hat{\mathbf{x}},\mathbf{u})+\hat{\mathbf{y}}$ by computing an additional differential equation (BLM). When the efficiency dominates, use $\mathbf{z}=\mathbf{h}(\hat{\mathbf{x}},\mathbf{u})$ suffering the inaccuracy only within $(0,T)$. 


\begin{figure}[tb!]
	\centering
	\includegraphics[width=0.98\columnwidth]{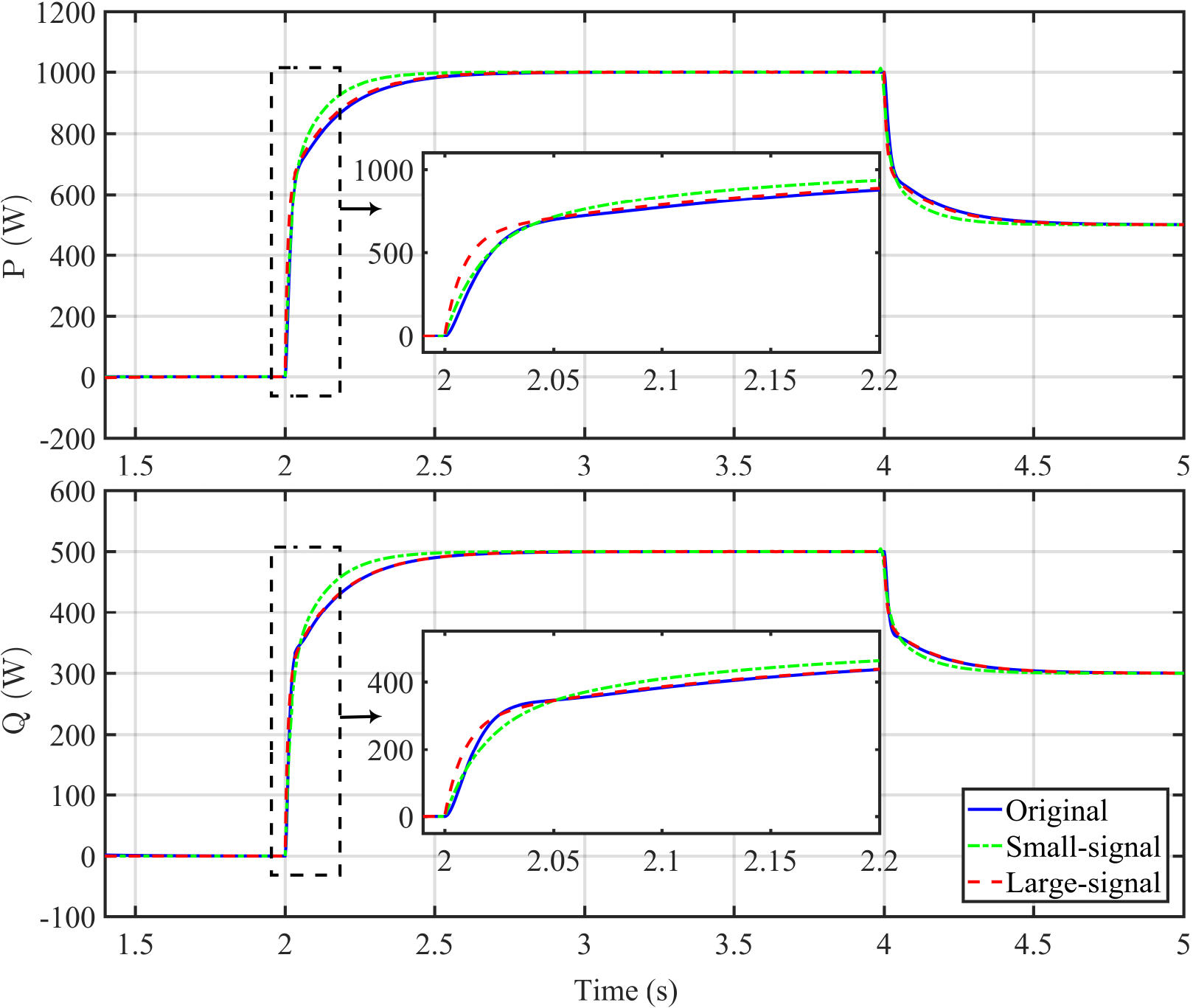}
	\caption{Simulation results of slow and fast dynamic responses of interested bus: active and reactive power.}
	\label{MGcompare1}
\end{figure}
\begin{figure}[tb!]
	\centering
	\includegraphics[width=0.98\columnwidth]{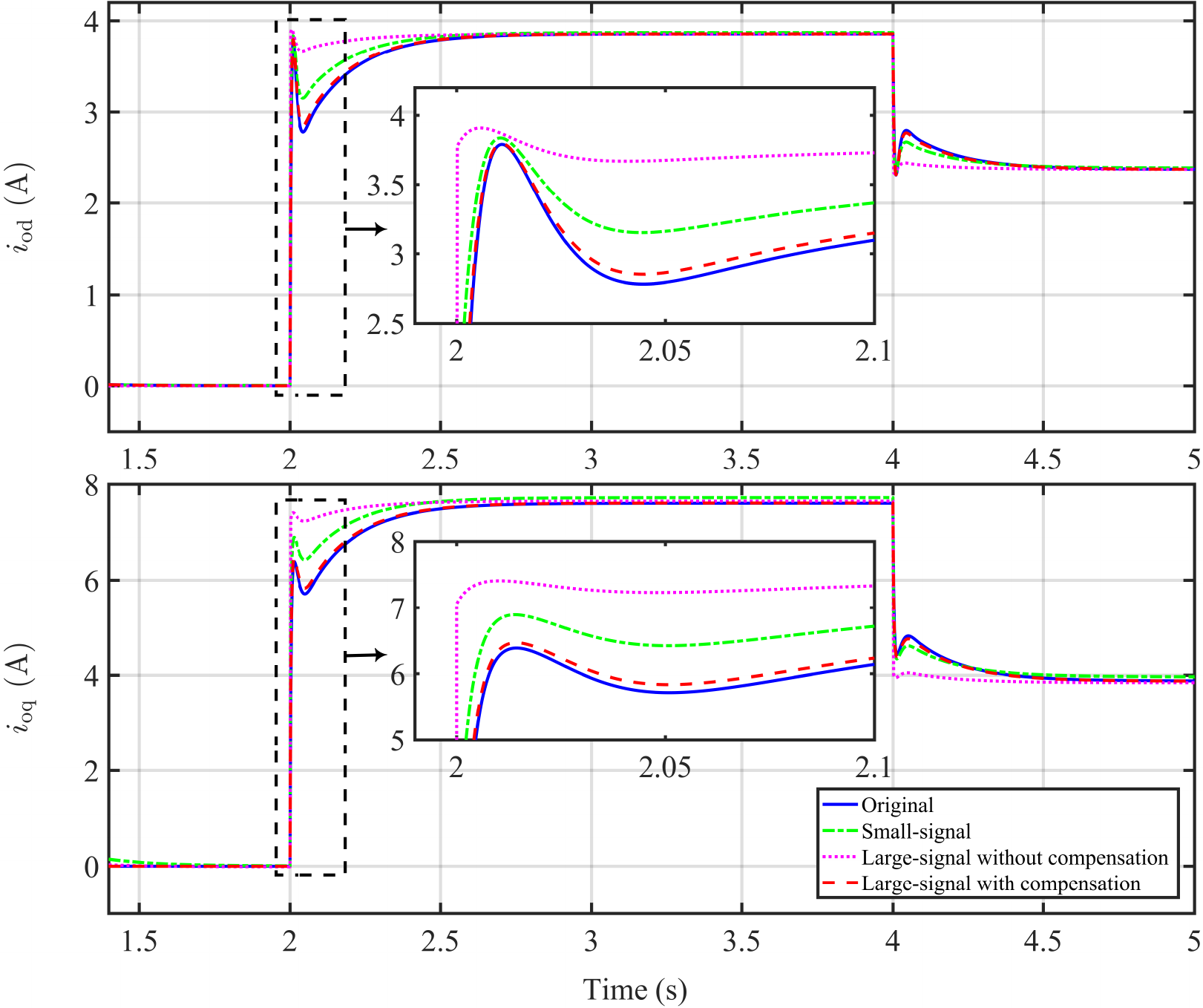}
	\caption{Simulation results of slow and fast dynamic responses of interested bus: $dq$-axis output currents $i_{\rm od}$ and $i_{\rm oq}$}
	\label{MGcompare2}
\end{figure}
\begin{figure}[tb!]
	\centering
	\includegraphics[width=0.98\columnwidth]{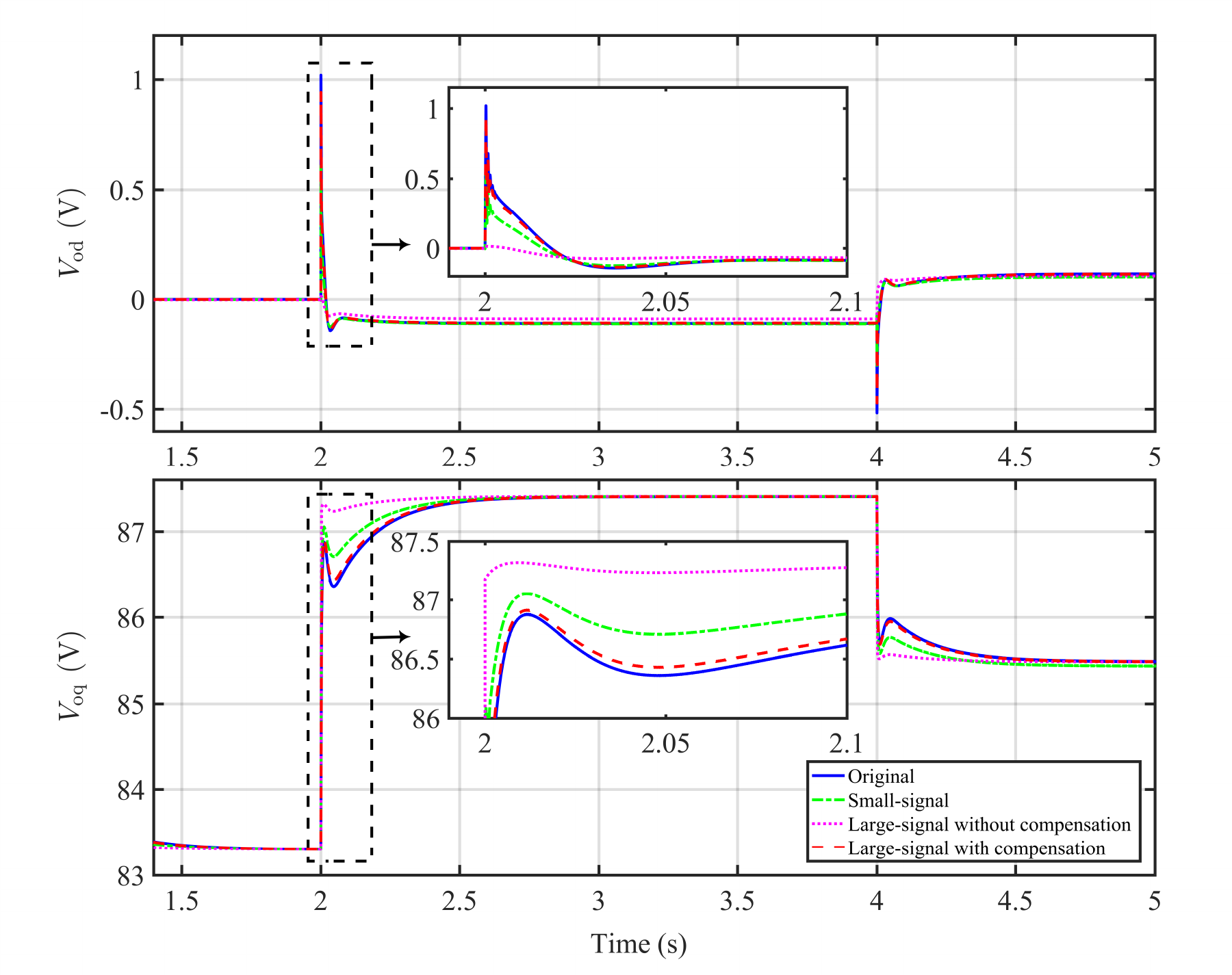}
	\caption{Simulation results of slow and fast dynamic responses of interested bus: $dq$-axis output voltages $V_{\rm od}$ and $V_{\rm oq}$.}
	\label{MGcompare3}
\end{figure}

Then we conduct the simulation of the derived ROM using Matlab. The active power command changes to $1000$ W at $2$ s and changes to $500$ W at $4$ s. The reactive power command changes to $500$ W at $2$ s and changes to $300$ W at $4$ s. A comparison simulation using the small-signal order reduction method in \cite{Rasheduzzaman2015} is conducted under the same conditions. The simulation results are shown in Fig. \ref{MGcompare1}-\ref{MGcompare3}, where blue solid lines denote the responses of the original model, green dash-dotted lines denote that using small-signal order reduction method, pink dotted lines denote the results of proposed LSOR without BLM compensation (i.e., QSS solution), and red dashed lines are the responses with the addition of solution $\hat{\mathbf{y}}$ of BLM (i.e. $\mathbf{z}=\mathbf{h}+\hat{\mathbf{y}}$). For the main slow dynamics (active and reactive powers) shown in (a), the proposed LSOR method is more accurate than the small-signal model during the transient period. As for the main fast dynamics voltages and currents shown in (b) and (c), the result of LSOR with compensation $\hat{\mathbf{y}}$ gives the most precise performance. However, the LSOR without $\hat{\mathbf{y}}$ gives worse performance than the small-signal one used in \cite{Rasheduzzaman2015}. This is because the fast dynamics predicted by the method in \cite{Rasheduzzaman2015} are also compensated with a corrected response. From the stability point of view, the red lines in Fig. \ref{MGcompare1}-\ref{MGcompare3} show that, with bounded input power commands, both ROM and BLM are stable, which indicates that the original system is stable as justified by the stability of blue lines.

In order to evaluate the computational performance, two different ordinary differential equation (ODE) solvers are adopted: ode45 solver and ode15s solver. The ode45 solver uses the 4th-order Runge-Kutta method with variable step sizes in order to solve the \textit{non-stiff} ODE problems, whereas the ode15s solver is designed for \textit{stiff} problems. The computational time is shown in Table \ref{time1}. This comparison indicates that our LSOR method converts the original model from a \textit{stiff} ODE problem to a \textit{non-stiff} one. The adoption of the proposed method also improves the stability of the ODE-solving process through this conversion. In conclusion, the proposed method can reduce the computational time from two aspects: the order of the system and the stiffness of the ODE problem.
\renewcommand\arraystretch{1.8}
	\begin{table}[tb!]
		\caption{Computational time of original and reduced-order models using different ODE solvers in grid-tied mode.}
		\setlength{\tabcolsep}{7pt}
		\begin{tabular}{|c|c|c|c|}	
			\hline
			\diagbox{Solver}{Model}&Original model&Reduced model&Percentage\\
			\hline
			ode45&$94.25$ s &$11.92$ s  & $87.4\%$ \\
			\cline{1-4}
			ode15s&$11.43$ s& $10.81$ s  &$5\%$ \\
			\hline
		\end{tabular}
		\label{time1}
	\end{table}

\emph{Remark 4:} Note that with the addition of the solution of BLM, we need to solve another set of differential equations. This seems that the proposed method has limited ability to reduce the computational burden. However, this is not the case. As discussed above, SPT reduces the computational burden not only by reducing the number of differential equations but also by converting the \textit{stiff} problem to a \textit{non-stiff} one. Moreover, the adopted example is a possible worst case that the perturbation coefficients are not small enough. When $\varepsilon$ is sufficiently small, the converging time $T$ can be sufficiently small as well. Then we can directly use the algebraic equation to estimate the fast states.

\subsection{Performance in Islanded Mode under Large Disturbances}
In this subsection, a simulation in islanded mode is conducted to verify the effectiveness of the proposed method by showing the dynamic responses of the buses with DERs. To study the dynamic characteristics, a $20\; {\rm \Omega}$ load is connected parallel to bus $12$ at $2$ s and disconnected at $2.5$ s. Following the similar procedure in case 1, we can identify the slow and fast dynamics of this multi-bus system. Despite the different parameter settings of inverters, the relative magnitudes of derivative terms' coefficients still hold uniformly. That means we can obtain a uniform division of slow and fast dynamics. This fact is based on the nature of different components'  time scales. The slow and fast states are divided as follows,
\begin{align}
\mathbf{x}_2&=\left[P_{i},Q_{i}, \varPhi _{{\rm PLL}i},  \delta_i, \varPhi _{{\rm d}i}, \varPhi_{{\rm q}i}, \Gamma_{{\rm d}i},\Gamma_{{\rm q}i},\right]^T,\\
\mathbf{z}_2&=\left[V_{{\rm odf}i},  I_{{\rm ld}i},  I_{{\rm lq}i},  I_{{\rm od}i}, I_{{\rm oq}i}, V_{{\rm od}i}, V_{{\rm oq}i}\right]^T. 
\end{align}

\begin{figure}[tb!]
	\centering
	\includegraphics[width=0.98\columnwidth]{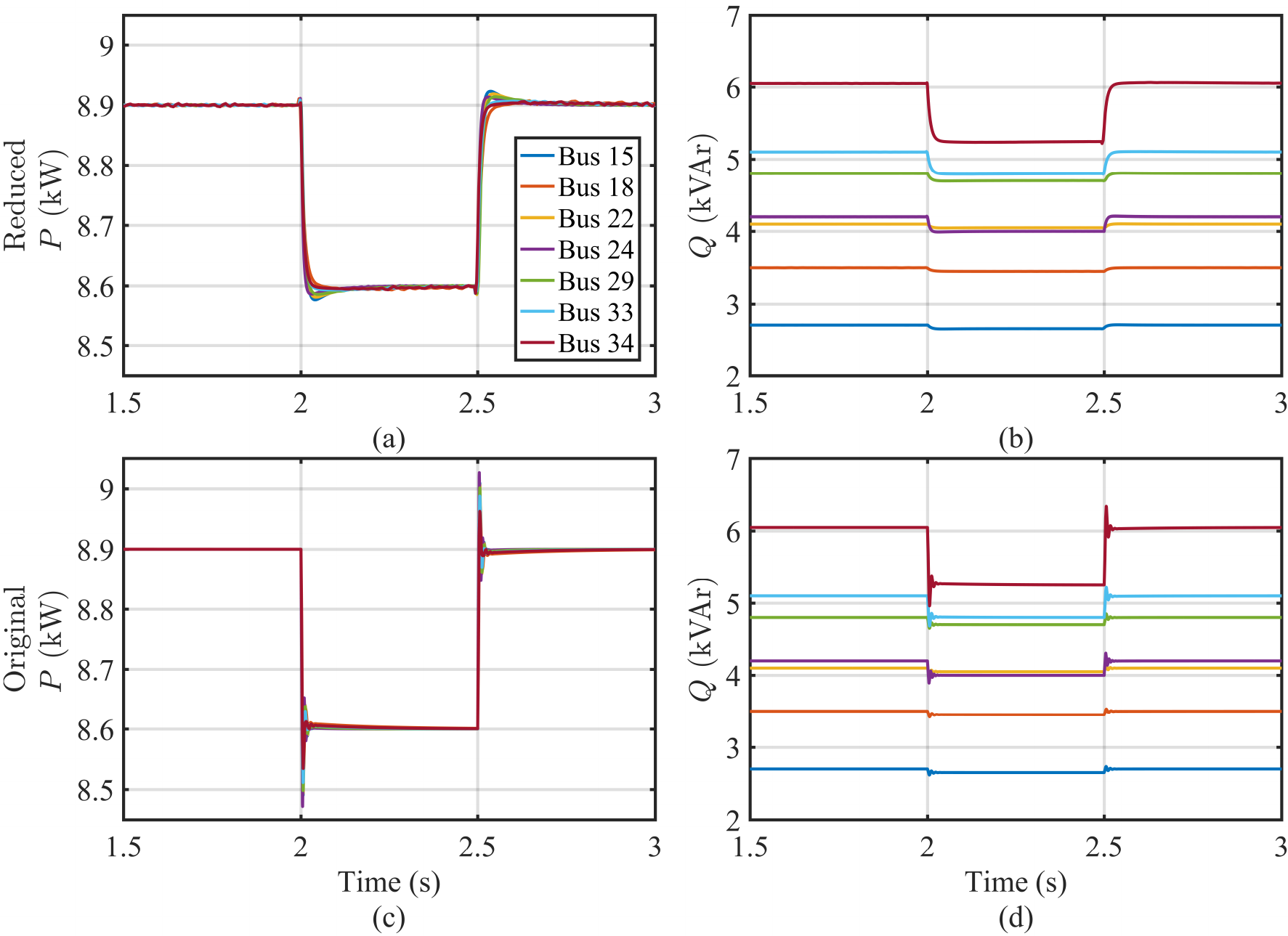}
	\caption{Comparison of the active/reactive power of the seven buses with DERs of original and reduced systems: (a)-(b) denote the responses of the reduced-order system, (c)-(d) are the responses of the original system.}
	\label{Multibus1}
\end{figure}
\begin{figure}[tb!]
	\centering
	\includegraphics[width=0.98\columnwidth]{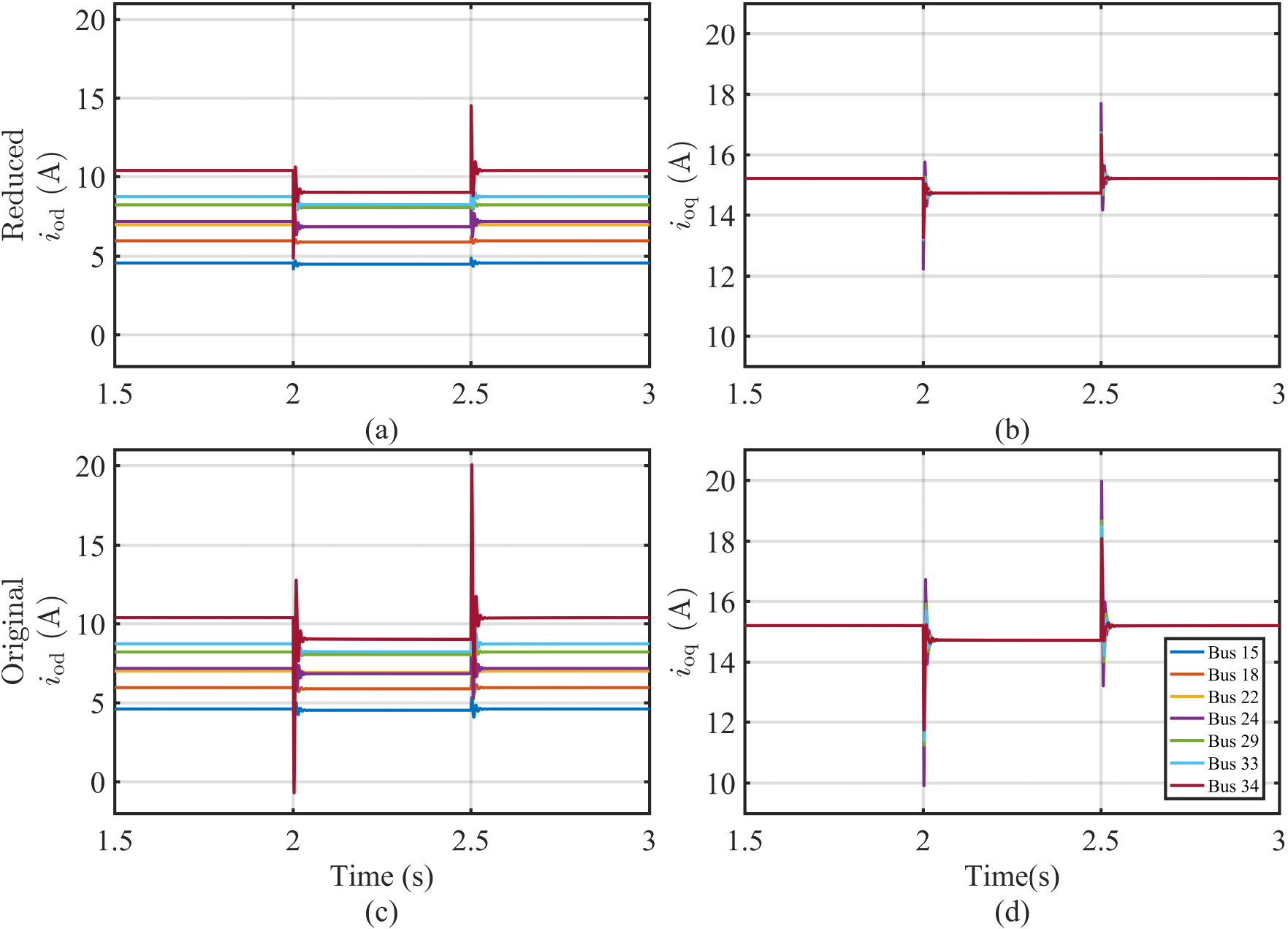}
	\caption{Comparison of the $dq$-axis output currents of the seven buses with DERs of original and reduced systems: (a)-(b) denote the responses of the reduced-order system, (c)-(d) are the responses of the original system.}
	\label{Multibus2}
\end{figure}
\begin{figure}[tb!]
	\centering
	\includegraphics[width=0.98\columnwidth]{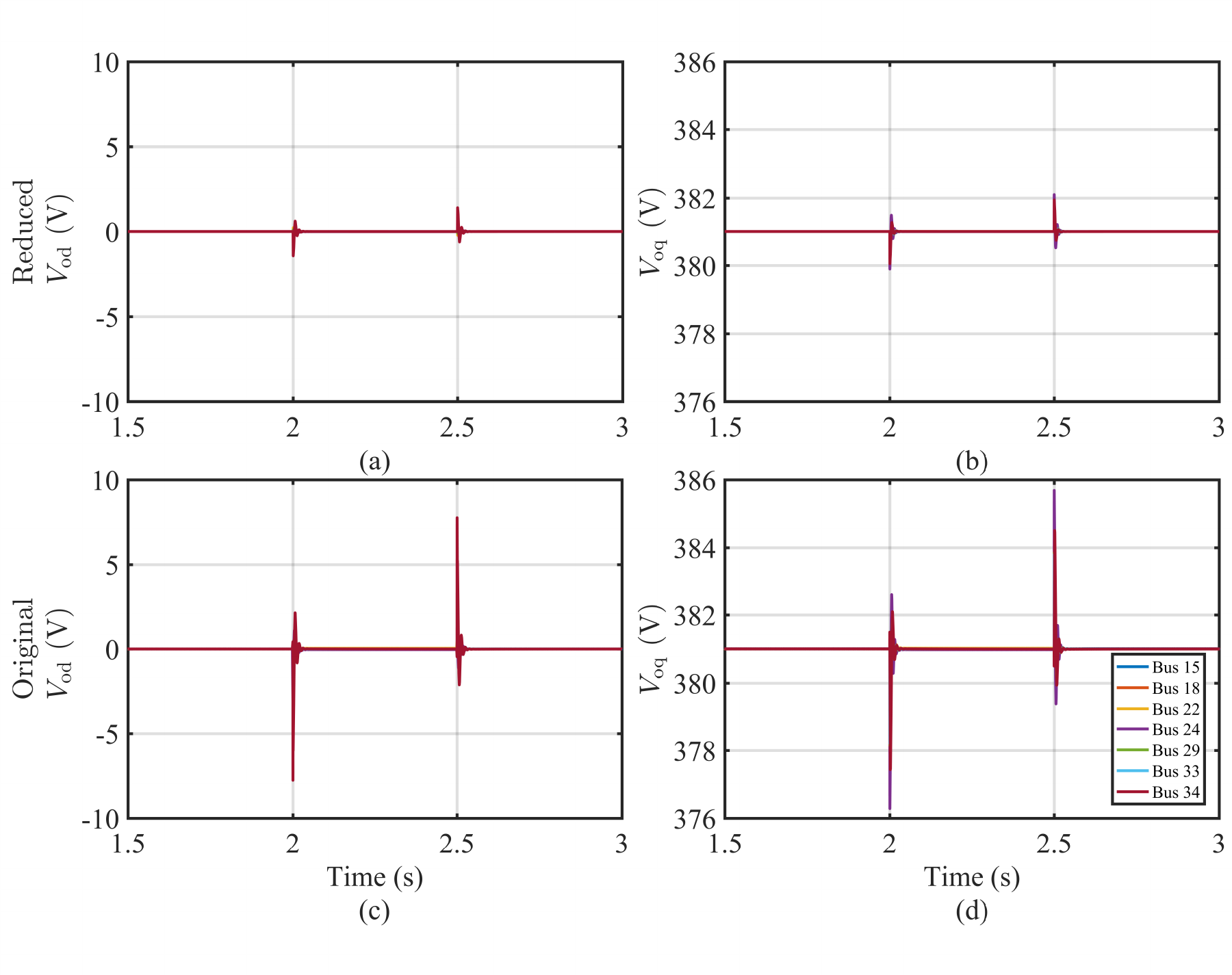}
	\caption{Comparison of the $dq$-axis output voltages of the seven buses with DERs of original and reduced systems: (a)-(b) denote the responses of the reduced-order system, (c)-(d) are the responses of the original system.}
	\label{Multibus3}
\end{figure}

The ROM can be derived using the \textit{Algorithm \ref{alg:LSOR}}. The order of the original model is reduced from $105^{\rm th}$ to $56^{\rm th}$. The simulation time is shown in Table \ref{time2}. Same as analyzed in the last case study, the proposed method can convert the stiff model of islanded MG to a non-stiff one to reduce the computational burden. Fig. \ref{Multibus1}-\ref{Multibus3} show the dynamic responses of the original and reduced models of seven buses with DERs. The comparison between the results of the original model and the reduced one shows the accuracy of the ROM. In addition, the responses under load sudden change verify the effectiveness of our method against large disturbances in islanded systems.
\renewcommand\arraystretch{1.8}
	\begin{table}[htb!]
		\caption{Computational time of original and reduced-order models using different ODE solvers in islanded mode.}
		\label{table4}
		\setlength{\tabcolsep}{7pt}
		\begin{tabular}{|c|c|c|c|}	
			\hline
			\diagbox{Solver}{Model}&Original model&Reduced model&Percentage\\
			\hline
			ode45&$104.25$ s &$11.25$ s  & $89.2\%$ \\
			\cline{1-4}
			ode15s&$13.23$ s& $11.37$ s  &$14\%$\\
			\hline
		\end{tabular}
		\label{time2}
	\end{table}
\section{Conclusion}\label{C5}
This paper proposes a large-signal order reduction (LSOR) approach for MGs in the electromagnetic transient (EMT) time scale with consideration of external control input by synthesizing a novel stability and accuracy assessment theorem. The advantages of our proposed theorem can be summarized into two aspects. Firstly, one can determine the stability of a full-order system by only analyzing the stability of its derived reduced-order model (ROM) and boundary layer model (BLM). Specially, when the ROM is input-to-state stable and the BLM is uniformly globally asymptotically stable, the original MGs system can be proved to be stable under several common growth conditions. This makes it easier and more feasible to determine the stability of a high-order system. Secondly, a set of quantitative accuracy assessment criteria is developed and embedded into a tailored feedback mechanism to guarantee the accuracy of the derived ROM. It is proved that the errors between solutions of reduced and original models are bounded and convergent under such conditions. The above stability and accuracy theorem has been strictly proven indicating that the proposed method is generic for arbitrary dynamic systems satisfying the given assumptions. Finally, we have conducted multiple simulations under different conditions on an IEEE standard MG system to verify the effectiveness of the proposed method.

\bibliographystyle{IEEEtran}
\bibliography{jiyoulishu}

\end{document}